\documentclass[sn-mathphys,Numbered]{sn-jnl}

\usepackage{graphicx}%
\usepackage{multirow}%
\usepackage{amsmath,amssymb,amsfonts}%
\usepackage{amsthm}%
\usepackage{mathrsfs}%
\usepackage[title]{appendix}%
\usepackage{xcolor}%
\usepackage{textcomp}%
\usepackage{manyfoot}%
\usepackage{booktabs}%
\usepackage{algorithm}%
\usepackage{algorithmicx}%
\usepackage{algpseudocode}%
\usepackage{listings}%

\theoremstyle{thmstyleone}%
\theoremstyle{thmstyletwo}%

\theoremstyle{thmstylethree}%

\begin{document}

\title{Perspective on ``Active Brownian Particles Moving in a Random Lorentz Gas''}

\author[1]{\fnm{C.} \sur{Reichhardt}}\email{reichhardt@lanl.gov}
\author*[1]{\fnm{C. J. O.} \sur{Reichhardt}}\email{cjrx@lanl.gov}

\affil*[1]{\orgdiv{Theoretical Division and Center for Nonlinear Studies}, \orgname{Los Alamos National Laboratory}, \orgaddress{\city{Los Alamos}, \state{New Mexico} \postcode{87545}, \country{USA}}}

\abstract{Self-propelled active matter can exhibit vastly
different behavior than systems with purely Brownian motion. 
In Eur.~Phys.~J.~E {\bf 40}, 23 (2017),
Zeitz, Wolf, and Stark
compared an active matter particle with a Brownian particle
moving in a random obstacle array.
They showed that near the obstacle percolation density, 
both Brownian and active particles exhibit the same
subdiffusive behavior, but the active particle
reaches a steady state more rapidly.
They also found that for high activity, the
active particle has a lower effective diffusion than the
Brownian particle due to the increased
self-trapping effect generated by the activity.
This result opens new directions for the
study of active matter in disordered media, including
bacteria in porous media,
active colloids on quenched disorder,
and active particles in crowded environments.
}
\maketitle

\section{Introduction}

\begin{figure}
\centering
\includegraphics[width=\columnwidth]{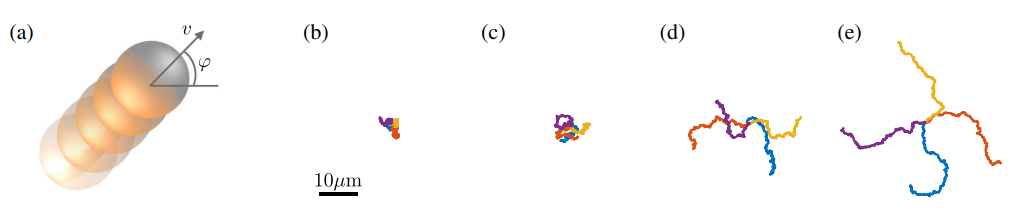}
\caption{(a) Illustration of a two-dimensional (2D) active Brownian particle
moving with speed $v$ at an orientation of $\phi$.
(b-e) Four representative 10 second particle trajectories for
a particle with radius $R=1 \mu$m and viscosity $\eta=0.001$ Pa s in water
at different velocities of
(b) $v=0 \mu$m/s (Brownian limit), (c) $v=1 \mu$m/s,
(d) $v=2 \mu$m/s, and (e) $v=3 \mu$m/s.
From Fig.~2 of \cite{Bechinger16}.}
\label{fig:1}
\end{figure}

Active matter particles break detailed balance via some form
of self-propulsion. The activity can take the form
of run-and-tumble dynamics, or
the particles can undergo driven diffusion
in which the particle velocity remains constant but
the direction of motion gradually changes over a characteristic
persistence length \cite{Marchetti13,Zottl16,Bechinger16}. 
The motion of a randomly moving particle can be characterized
by its mean-square displacement over time,
$\langle r^2(t)\rangle \propto D_\alpha t^\alpha$,
where $D_\alpha$ is the generalized diffusion constant and
$\alpha$ is the diffusive exponent.
For a Brownian particle, $\alpha = 1.0$, but if the
motion of the particle is slowed by trapping or crowding,
then $0 < \alpha < 1.0$
and the system is said to exhibit subdiffusion \cite{Bouchaud90}.
In the case of ballistic motion, $\alpha = 2.0$,
and if collisions occur
that cause the ballistically moving particle to change direction,
superdiffusive behavior emerges for which $1.0 < \alpha \leq 2.0$.
An active particle generally undergoes
superdiffusive motion on short or intermediate times,
but at long times, its motion becomes diffusive due to the ever-changing
direction of travel
\cite{Marchetti13,Bechinger16}. 
Figure~\ref{fig:1} shows an example of a
driven diffusive or active Brownian particle moving in
two dimensions (2D).
The particle has a propulsion velocity of $v$ and an orientation
direction $\phi$ that gradually changes over time.
Example trajectories for $v$ values ranging from the Brownian
limit of $v=0 \mu$m/s to $v=3 \mu$m/s 
appear in Figs.~\ref{fig:1}(b-e). 
As $v$ increases, the trajectories are able to extend further from
the $t=0$ position of the particle before the
direction of motion becomes
fully randomized.
Both single active particles and assemblies of active particles
have been extensively studied,
but there are additionally
numerous situations in which
active particles can interact with some
form of barrier or obstacle
\cite{Chepizhko13,Reichhardt14,Bechinger16,Khatami16,Morin17,Volpe17,Reichhardt17a,Reichhardt18a,Bertrand18,Duan21,Rizkallah22,Brady22,Saintillan23,Pietrangeli25}.
One example of such a system is a bacterium in a disordered medium \cite{Bhattacharjee19a,Irani22,Datta25,Mattingly25}.
A natural question is how the motion of an active particle in
a disordered environment, such as an assembly of obstacles,
would differ from that of a Brownian particle.
While the activity could lead to higher diffusion
and larger displacements
compared to a Brownian particle due to the
correlated propulsion,
the persistence in motion could also result in a self-trapping effect. 

\begin{figure}
\centering
\includegraphics[width=0.8\columnwidth,trim=140 120 140 100,clip]{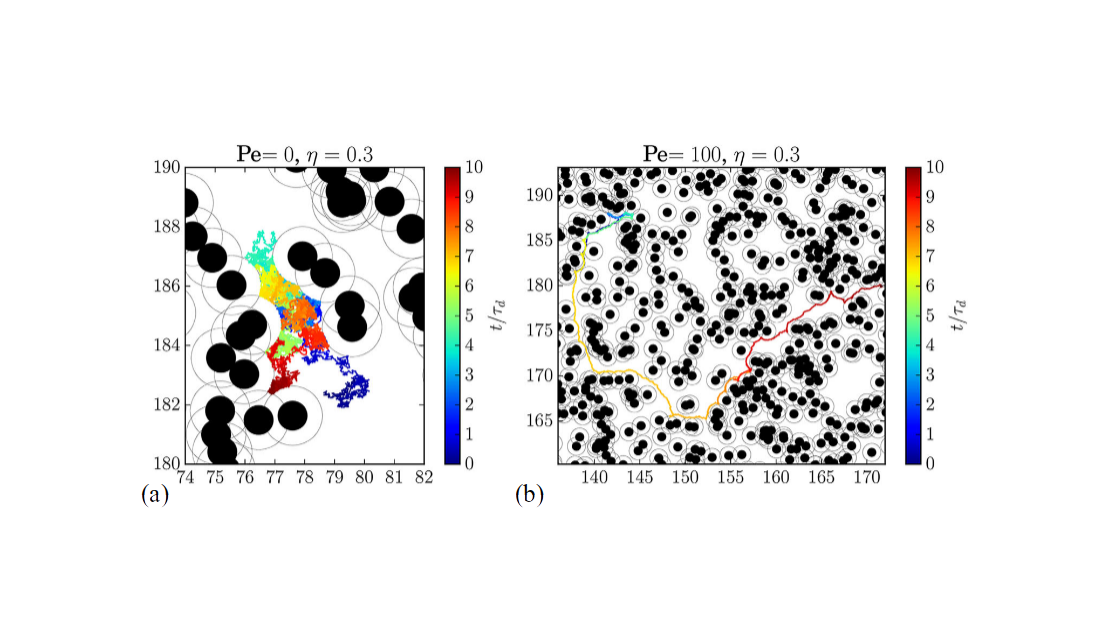}
\caption{Particle trajectories with time $t/\tau_d$ encoded by color
for (a) a Brownian particle and (b) an active Brownian particle in a
system with a background obstacle density of $\eta=0.3$.  
From Fig.~1 of \cite{Zeitz17}.}
\label{fig:2}
\end{figure}

Zeitz {\it et al.} \cite{Zeitz17} considered an active
particle moving through a 
random obstacle array, which is also known as a Lorentz gas \cite{Beijeren82}.
A system of area $A$ can be described in terms of
the reduced density or the area covered by the disk-shaped obstacles,
$\eta = \pi R^2/A$, where $R$ is the disk radius.
A percolation transition of the obstacles occurs
near $\phi = 0.67$ \cite{Torquato02} in 2D.
Zeitz {\it et al.} focused on
active and Brownian particles of finite size, such that
the critical percolation occurs at a reduced density of
$\eta=0.28$, and measured the mean-square particle displacements
and effective diffusion constants as the
obstacle density was swept upward through the percolation transition.
For the Brownian particles,
the motion is diffusive at lower $\eta$
values well below the percolation density,
but close to the percolation density,
the particles become increasingly trapped
and the motion becomes sub-diffusive, similar to
what is observed in a glass.
Figure~\ref{fig:2}(a) shows the time-encoded
trajectory of a Brownian particle moving through an obstacle array
at $\eta=0.3$. For an active particle in the same array,
Fig.~\ref{fig:2}(b) indicates that
there are extended regions where
the particle moves ballistically between obstacles.

\begin{figure}
\centering
\includegraphics[width=0.8\columnwidth,trim=150 30 90 0,clip]{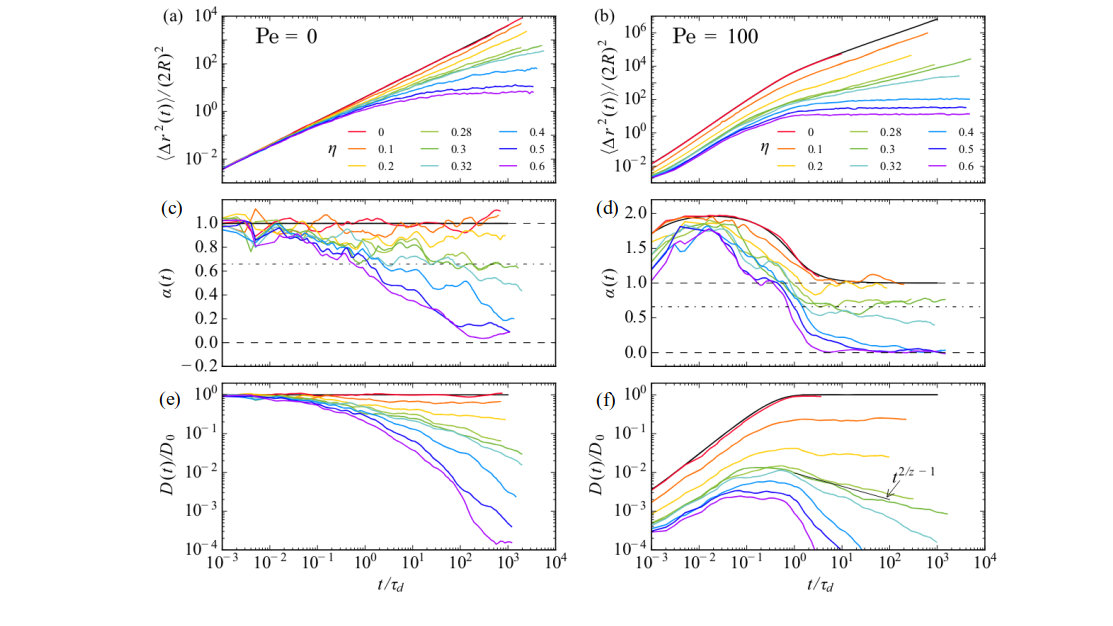}
\caption{(a, b) Mean squared displacement
$\langle \Delta r^2(t)\rangle/(2R)^2$, (c, d)
local exponent $\alpha(t)$, and (e, f)
local diffusion coefficient $D(t)/D_0$ versus
scaled time $t/\tau_d$ for (a, c, e) a Brownian particle with $Pe=0.0$ and
(b, d, f) an active Brownian particle with $Pe=100$ at obstacle
densities ranging from $\eta=0.0$ to $\eta=0.6$.
The percolation transition occurs near $\eta_c=0.28$.
The active particles have an
extended region at short times where the motion is superdiffusive,
but for $\eta > \eta_c$ the motion of the active particles is reduced
compared to the Brownian particles due to self trapping effects.
From Fig.~2 of \cite{Zeitz17}.}
\label{fig:3}
\end{figure}

The motion of the Brownian particle is diffusive at low obstacle
density but becomes subdiffusive
as $\eta$ approaches the critical percolation threshold.
A key advantage to the Lorentz gas model is that its
percolation universality class is known,
which gives a subdiffusive exponent of $\eta_c \approx 0.66$.
Figure~\ref{fig:3}(a) shows the
mean-square displacement versus time for a Brownian
particle at varied obstacle densities from
$\eta=0.0$ to $\eta=0.6$.
The activity level is characterized by the P{\' e}clet number ($Pe$), with
large $Pe$ indicating high activity, so that
Brownian particles have $Pe = 0.0$.
The local diffusion exponent
$\alpha(t) = d\log(\langle\Delta r^2(t)\rangle)/d\log(t)$
and the local effective diffusion constant
$D(t)/D_0$ are shown in Fig.~\ref{fig:3}(c) and (e),
respectively.
For $\eta < \eta_c$, $\alpha(t) = 1.0$, indicating
that regular diffusion is occurring.
As $\eta$ approaches
the critical $\eta_c$, $\alpha(t)$ approaches 0.66, indicated
by the dashed lines in Fig.~\ref{fig:3}(c), while when $\eta>\eta_c$,
$\alpha(t)$ approaches 0.0 and the diffusion constant monotonically decreases
with time.
Figure~\ref{fig:2}(b) shows the
mean squared displacements versus time for the same system but with an active
particle of $Pe = 100$.
In Fig.~\ref{fig:2}(d), the corresponding
local exponent obeys $\alpha(t) > 1.0$,
indicative of superdiffusion, as expected for an active particle.
In the $\eta=0$ system where there are no obstacles,
the system shows superdiffusion at short times
but has a crossover to regular diffusion at longer times.
Near the critical density $\eta \approx \eta_c$,
the active particles have an extended region of time in which
$\alpha = 0.66$, indicating that the active particles
are able to reach a steady state more rapidly than
the Brownian particles.
For $\eta > \eta_c$,
the active particles eventually reach $\alpha(t) = 0.0$,
and the decrease in $D$ at long times is much more rapid than
for the Brownian particles.
These results indicate that when $\eta > \eta_c$,
the active particles are actually less motile than the
Brownian particles as the result of a
self-trapping effect in which the 
active particle can become trapped behind an obstacle.
The persistence of the active motion holds the active particle in
the trapping site until the velocity has rotated
far enough into a new direction to transport the active
particle away from the trapping obstacle; however,
the particle rapidly becomes trapped behind a different
obstacle.
Brownian particles constantly change their direction of motion and
are therefore better able to
explore the available free space compared to the active particles.
Self-trapping of active particles has been observed in
other active matter systems, including run-and-tumble systems, where
for a given amount of substrate disorder there is often
an optimal activity level that maximizes the motility of the active
particles
\cite{Reichhardt14,Volpe17,Bertrand18,Rizkallah22,Saintillan23,Mattingly25}.

\begin{figure}
\centering  
\includegraphics[width=0.8\columnwidth,trim=130 260 170 100,clip]{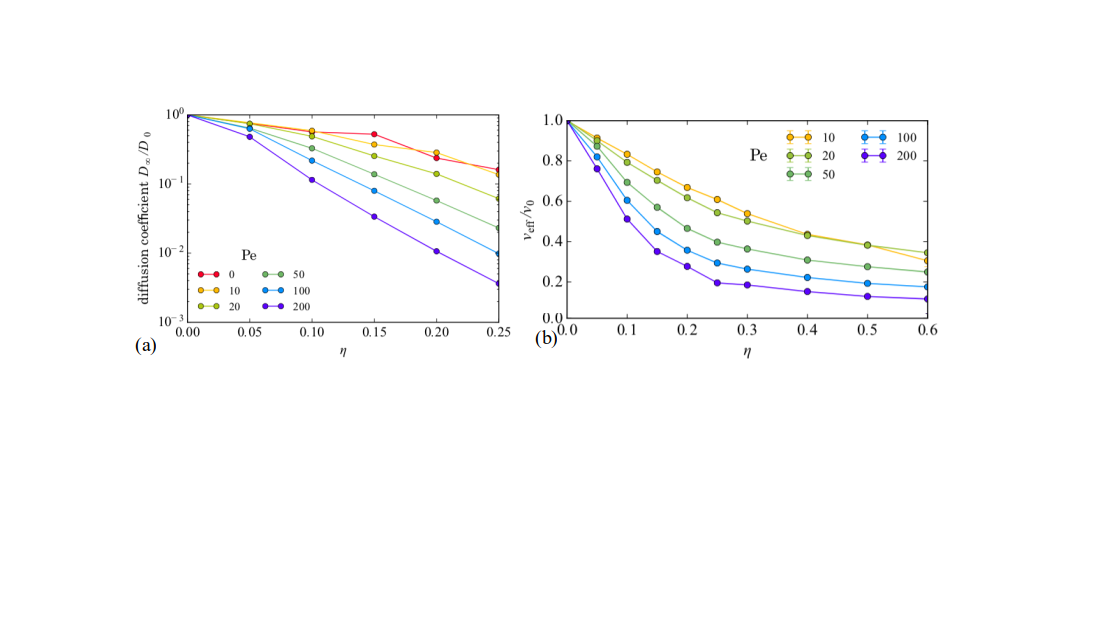}
\caption{Diffusive and ballistic motions at different P{\' eclet}
values. (a) The diffusion coefficient $D_\infty/D_0$ vs $\eta$, where
$D_0$ is the $\eta=0$ diffusion coefficient.
From Fig.~5 of Ref.~\cite{Zeitz17}.
(b) The effective propulsion velocity $v_{\rm eff}/v_0$ vs $\eta$, where
$v_0$ is the propulsion speed.
From Fig.~7(a) of Ref.~\cite{Zeitz17}.}
\label{fig:4}
\end{figure}

Figure~\ref{fig:4}(a) shows the long-time diffusion
$D_{\infty}/D_0$ versus $\eta$ measured in the range $\eta<\eta_c$
for varied $Pe$ \cite{Zeitz17}.
Here, $D_0$ is the obstacle-free diffusion constant.
For $Pe = 0.0$, the diffusion decreases by
a factor of less than 10 as $\eta$ increases.
In contrast, for $Pe = 200$, the diffusion drops by
300 times, illustrating the decrease in mobility caused by introducing
activity.
This behavior is underscored by the plots of the effective propulsion
velocity $\eta/v_0$ in Fig.~\ref{fig:4}(b), where
the velocity is higher
at lower $Pe$ where the particles are able to move through the
free spaces of the obstacle assembly,
whereas particles with large $Pe$ spend a significant fraction of time
immobilized in trapping sites behind obstacles.

\begin{figure}
  \begin{minipage}{\columnwidth}
  \centering
    \includegraphics[width=0.8\columnwidth]{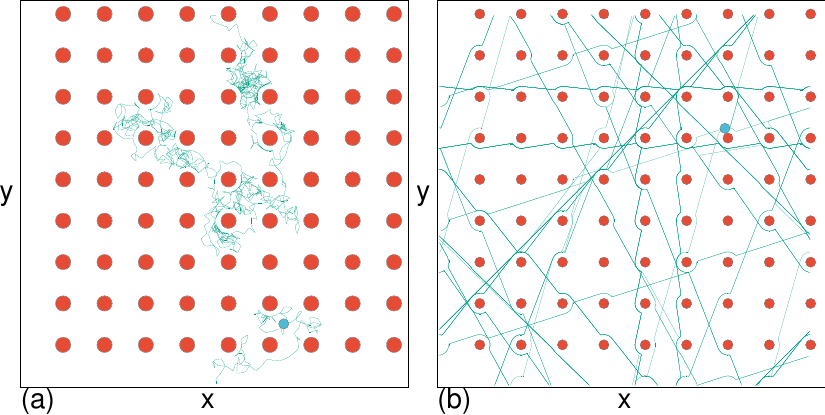}
  \end{minipage}\hfill
  \begin{minipage}{\columnwidth}
  \centering
    \includegraphics[width=0.8\columnwidth]{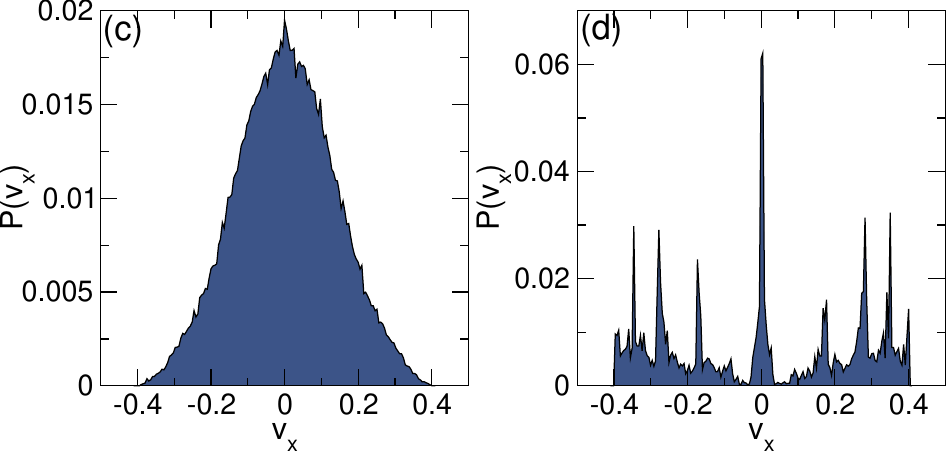}
  \end{minipage}\hfill
\caption{(a, b) Motion of individual particles through a
periodic array of posts (red) with lattice constant $a_s$.
(a) Trajectory of a Brownian particle.
(b) Trajectory of a run-and-tumble active particle
with run correlation length $l_a=20 a_s$. The motion is channeled
along substrate symmetry directions.
(c, d) The corresponding distribution of instantaneous $x$ direction
velocities $P(v_x)$ for (c) the Brownian particle from panel (a), where
the distribution is Gaussian, and (d) the active particle from
panel (b), where the peaks indicate the locking of the motion to
the substrate symmetry directions.
From Fig.~\ref{fig:2} of \cite{Reichhardt22a}.
	}
\label{fig:5}
\end{figure}

A number of studies have built upon the work of
Zeitz {\it et al.}
For example, the active particles
can move through periodic arrays of obstacles \cite{Volpe11,Reichhardt20,Reichhardt21a,Dehkharghani23}.
A Brownian particle traversing a periodic obstacle array, as illustrated
in Fig.~\ref{fig:5}(a), diffuses between the obstacles and
has the Gaussian velocity distribution $P(v_x)$ shown in
Fig.~\ref{fig:5}(c).
The case of a run-and-tumble active particle is shown in
Fig.~\ref{fig:5}(b). Here the particle travels ballistically for
long distances between the obstacles along preferred symmetry
directions of $\theta_m=0^\circ$, 45$^\circ$, and $90^\circ$.
As the radii of the obstacles decreases, the number of available
easy flow directions increases, and for the square lattice shown in
the figure,
these correspond to motion
along $\theta_m= \tan^{-1}(n/m)$, where $n$ and $m$ are integers.
The active particle motion remains superdiffusive out to longer times
compared to a system with randomly placed obstacles,
and the velocity distribution $P(v_x)$ is
highly non-Gaussian
and contains peaks associated
with motion along the easy flow directions of the substrate,
as shown in Fig.~\ref{fig:5}(d) \cite{Reichhardt20}. 

\begin{figure}
\centering  
\includegraphics[width=0.8\columnwidth,trim=260 120 200 70,clip]{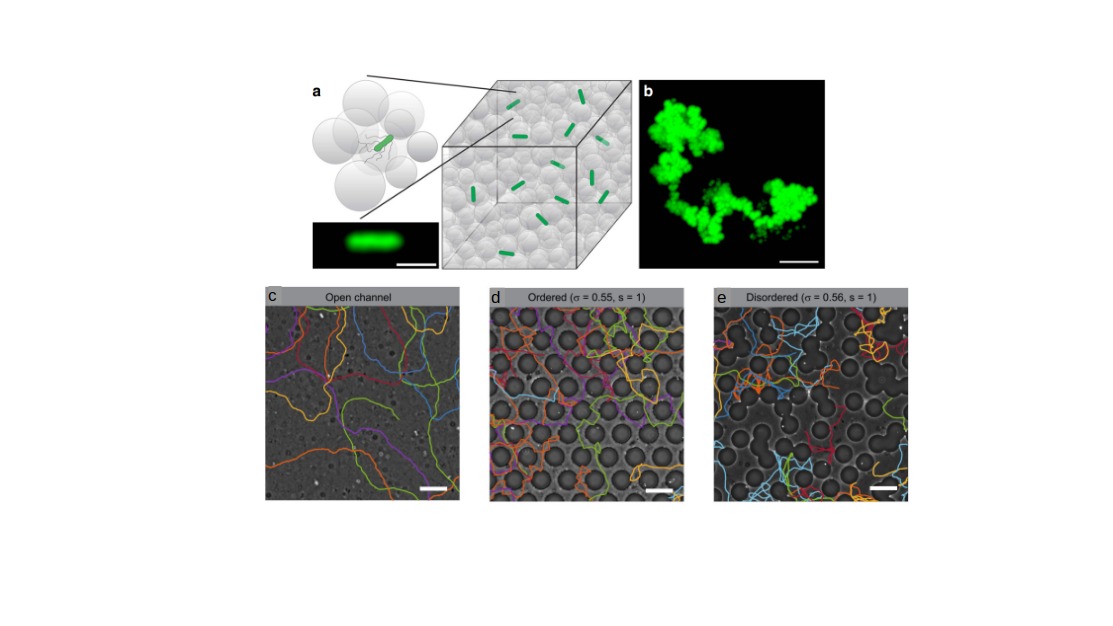}
\caption{(a,b) Bacteria moving in a three-dimensional (3D)
array of hydrogel spheres, shown (a) schematically and (b) from
fluorescent imaging of a tracer particle.
From Fig.~1 of \cite{Bhattacharjee19a}.
(c,d,e) Bacterial motion in (c) smooth,
(d) triangular, and (e) disordered obstacle arrays.
From Fig.~1(a,b,c) of \cite{Dehkharghani23}.}
\label{fig:6}
\end{figure}

There have now been numerous experimental realizations of the
Zeitz {\it et al.} model.
Figure~\ref{fig:6}(a,b) shows a three-dimensional (3D) hydrogel containing
tortuous paths for bacteria motion. Here, the bacteria exhibit intermittent
hopping from pore to pore of the medium, and there is
a crossover from superdiffusion at short times to
regular diffusion at long times, with 
the net diffusion decreasing
with increasing obstacle density \cite{Bhattacharjee19a}.
Figure 6(c,d,e) shows experimental trajectories
for bacteria moving in 2D
on a smooth substrate, a triangular
obstacle array, and a random obstacle array,
demonstrating different modes of motion.
In this case, for shorter times,
the motion on the ordered arrays shows locking along symmetry
directions of the substrate;
however, at longer times, the motion
for both ordered and random arrays
exhibits similar behavior \cite{Dehkharghani23}. 

\section{Future Directions}

As active matter becomes an increasingly mature field,
there will be more interest in studying
active matter systems with increased complexity
in terms of the rules of how the particles move,
interact with each other,
or transport themselves through random or structured environments.
Here, we highlight some possible future extensions for the Zeitz model:
(1) The active particle could change its behavior over time.
For example, if the active particle reaches
a state where it is no longer moving due to trapping,
it could decrease its P{\' e}clet number
for a fixed time in order to escape the trapping site.
Conversely, if the particle
is moving rapidly for extended periods,
it could increase its P{\' e}clet number to amplify the motion,
or it could reverse or otherwise modify its swimming direction.
In this way, the particle could optimize its motion via feedback.
(2) The spherical active particles could be replaced by
more complex particle geometries, such as polymers, rods,
flexible chains, or deformable active particles moving through obstacles.
There has already been some work treating
active polymers in disordered media \cite{Kurzthaler21}. 
(3) Variations of the model could be applied to the spreading of infections.
Some work has already been done on epidemic transmission through
active matter systems \cite{Forgacs22},
but it would be interesting to explore
the effective diffusion in an infectious model.
(4) The particles and/or the environment could be made chiral.
Some groups have considered
chiral active particles in obstacle arrays \cite{Chepizhko20,Lou22,Chan24}. 
These systems could have interesting topological effects or edge modes.
(5) Different substrate geometries could be introduced, such as disordered
hyperuniform, fractal, quasiperiodic, or anisotropic.
(6) Models could be considered
in which the obstacles themselves are dynamical
and can move, either independently or in response
to the motion of the active particles.
(7) Collective effects could be captured
by considering multiple interacting active particles.
(8) The active particles could move over different types of complex networks.

Many of the studies performed to date have focused
on diffusion and diffusion exponents,
but there are also many other quantities that
could be explored, such as aging effects and memory.
Often studies are performed on systems
where there is no external drive on the active particles;
however, it would be possible to
add a drift force, which could arise from a
fluid or an applied field.
In the case of
active particles in a fluid,
additional hydrodynamic effects can arise,
so in the presence of a porous medium,
the medium itself could induce hydrodynamic flows with which the
active particles could interact.
In this case, there could be hydrodynamic-induced trapping effects,
or alternatively the hydrodynamics might
cause the particles to avoid entering trapping sites.
In addition to applying dc flows or fields, the particles could be subjected
to ac driving, or to a combination of ac and dc driving,
which could increase the diffusion.

Many active matter studies related to the Zeitz model
have been performed in 2D,
but a number of real-world applications would be in 3D,
which would give a different percolation threshold.
Also, studies in both 2D and 3D have focused on
impenetrable or penetrable obstacles;
however, the substrate
could have other topological features,
as in the case of gels or clusters,
where there could be multiple percolation transitions
for different densities or where the percolation could occur anisotropically
and appear first in one direction and then in another.
Instead of non-penetrable obstacles,
the system could contain pinning sites,
localized changes in the effective friction,
or random landscapes where active particles would be able to
hop over higher energy barriers.
In addition to varying
the density of the obstacles, it would also be possible to change
the nature of the interaction of the particles with the obstacles.
A particle could move along a wall, reflect from it, be attracted by
it, or be repelled by it; it is also possible to modify
the roughness of the walls or introduce
increased or decreased friction for
a particle that is sliding or moving along a wall.
In some cases, the interaction with the walls could change
the propulsion mechanics, causing active particles to move faster
when in contact with a wall.
If the obstacles are able to move, additional percolation transitions
could occur, with one transition 
associated with contact percolation at a low density,
and another associated with a jamming transition at higher density.

\section{Summary}

Active matter is a rapidly growing interdisciplinary field with applications to biological systems, statistical physics, soft matter, social systems, and robotics. Early studies focused on single or interacting particles moving over a smooth substrate.
Zeitz {\it et al.} \cite{Zeitz17}
introduced a model for an active particle moving on an obstacle array
to create an active matter version of the Lorentz gas. Despite the
apparent simplicity of this model,
Zeitz {\it et al.}
found that it could exhibit a number of interesting effects,
such as showing the same behavior near percolation as Brownian particles.
Additionally, the activity can actually decrease the diffusion
in these systems by causing increased trapping.
The model of Zeitz {\it et al.} can be applied
to a wider range of active matter systems,
including run-and-tumble particles, active particles on periodic substrates,
and even deformable substrates,
and there have been a number of theoretical extensions
and experimental realizations of this system.
These include active colloids on patterned substrates,
swimming bacteria in 2D and 3D systems,
ordered and disordered substrates, active polymers,
and active motion on periodic substrates.
A growing number of new research directions
have been inspired by the work of Zeitz {\it et al.},
which will lead to
more studies of active matter in complex environments,
the development of controlled
modes of motion for active particles,
and the creation of increasingly complex environments.
The work of Zeitz {\it et al.} also nicely brings together
established ideas from statistical mechanics,
such as percolation, with new ideas in nonequilibrium systems and active matter.

\noindent{\textbf{\textsf{Data availability}}}\\
No datasets were generated during and/or analyzed during the current study.

\noindent{\textbf{\textsf{Acknowledgments}}}
We gratefully acknowledge the support of the U.S. Department of
Energy through the LANL/LDRD program for this work.
This work was supported by the US Department of Energy through
the Los Alamos National Laboratory.  Los Alamos National Laboratory is
operated by Triad National Security, LLC, for the National Nuclear Security
Administration of the U. S. Department of Energy (Contract No. 892333218NCA000001).

\noindent{\textbf{\textsf{Competing Interests}}}\\
The authors declare no competing interests.

\noindent{\textbf{\textsf{Author contributions}}}\\
All authors contributed equally to this work.

\includegraphics[width=0.35\columnwidth, trim=100 250 100 100, clip]{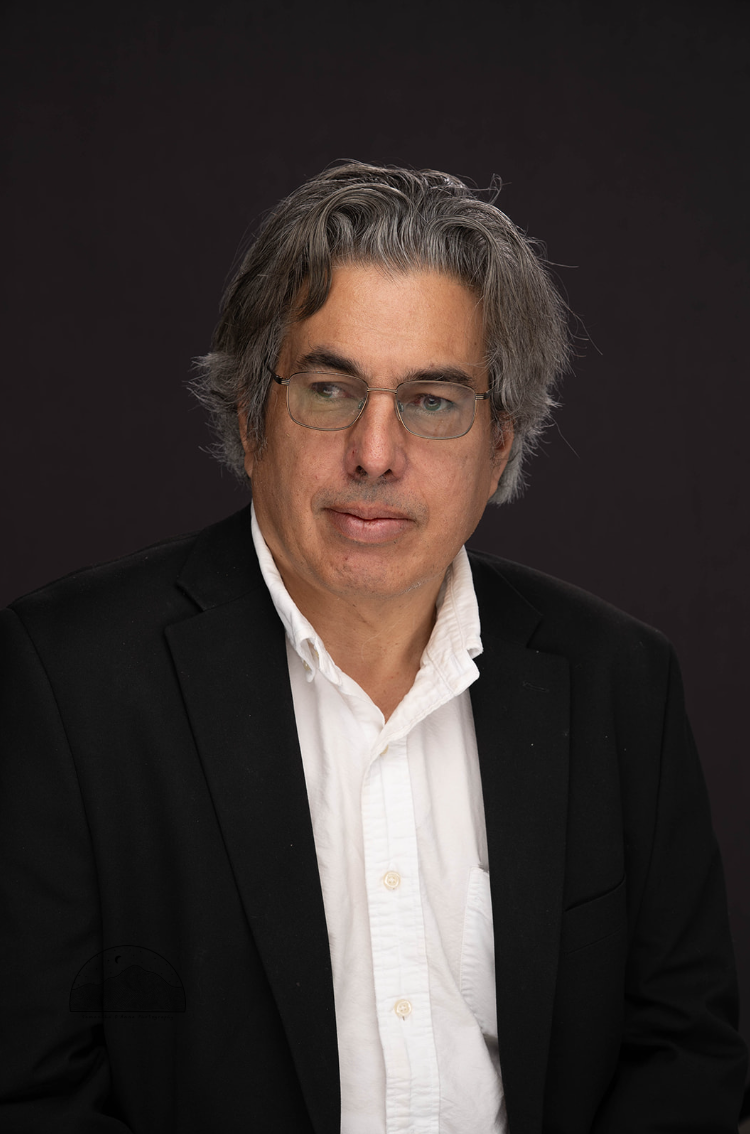}
\textbf{Charles Reichhardt}
is currently a Senior Scientist in the
Theoretical Division at Los Alamos National Laboratory (LANL). He
earned his PhD in condensed matter physics at the University of Michigan
in 1998. He was a postdoc at the University of California -- Davis and
a Feynman Fellow distinguished postdoc at LANL before becoming
a LANL staff member in 2003. He is an expert in large scale particle based simulations of nonequilibrium systems including superconducting vortex dynamics,
magnetic skyrmions, and active matter.
He was elected an American Physical Society Fellow in 2011 and
a Los Alamos National Laboratory Fellow in 2025.

\includegraphics[width=0.35\columnwidth, trim=50 70 10 20, clip]{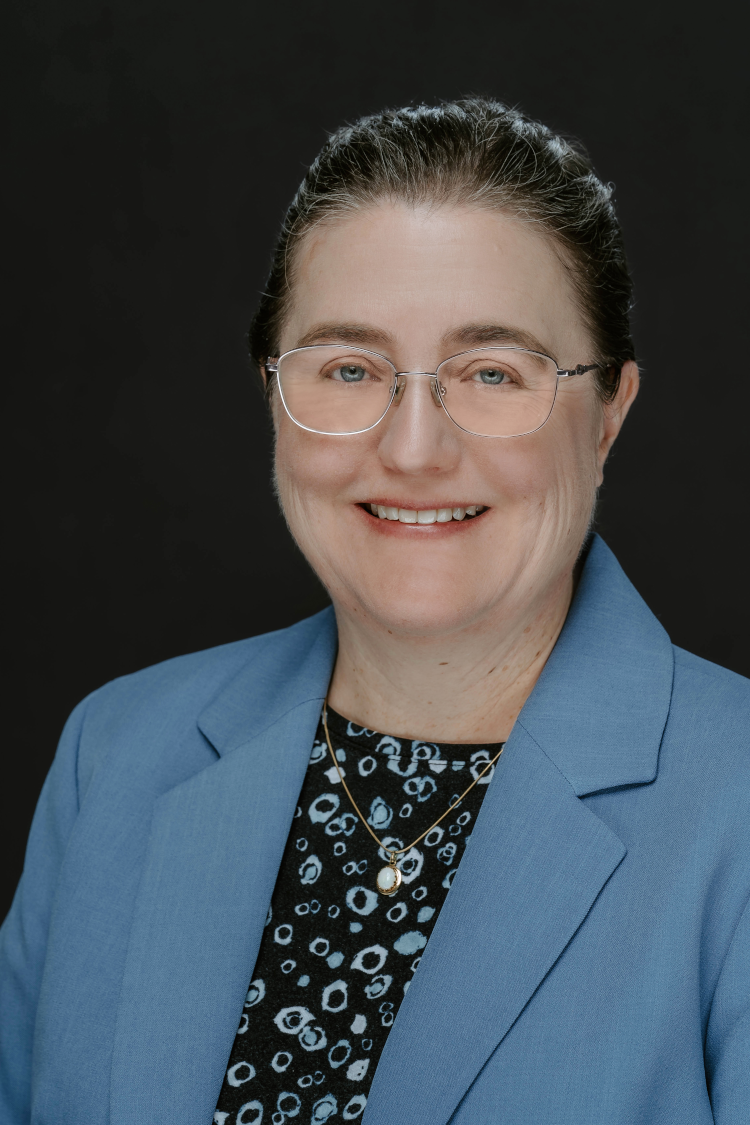}
\textbf{Cynthia Reichhardt}
is currently a Senior Scientist in the Theoretical Division at
Los Alamos National Laboratory (LANL). She received a PhD in
condensed matter physics
from the University of Michigan in 1998 before taking a postdoctoral
position at the University of California -- Davis. She became a
Director's Fellow distinguished postdoc at LANL before becoming a
LANL staff member in 2003.
She is an expert in computational simulations of nonequilibrium transport including superconducting vortex dynamics, active or self-propelled particles, magnetic skyrmions, granular matter, colloidal particles, ion damage, reaction-diffusion systems, aging, and corrosion.
She has been an American Physical Society (APS) fellow since 2011, served as Chair of the APS Topical Group on Statistical and Nonlinear Physics, received the
LANL Fellows' Prize for Outstanding Research in Science or Engineering in 2018,
and was elected a Los Alamos National Laboratory Fellow in 2024.

\bibliography{mybib}

\end{document}